**Atomistic determination of flexoelectric properties of crystalline dielectrics**

R. Maranganti[1] and P. Sharma[1,2,♣]
[1]Department of Mechanical Engineering,
[2]Department of Physics
University of Houston, Houston, TX, 77204, U.S.A

**Abstract:** Upon application of a uniform strain, internal sub-lattice shifts within the unit cell of a *non-centrosymmetric* dielectric crystal result in the appearance of a net dipole moment: a phenomenon well known as piezoelectricity. A macroscopic *strain gradient* on the other hand can induce polarization in dielectrics of any crystal structure, even those which possess a centrosymmetric lattice. This phenomenon, called *flexoelectricity*, has both bulk and surface contributions: the strength of the bulk contribution can be characterized by means of a material property tensor called the *bulk flexoelectric tensor*. Several recent studies suggest that strain-gradient induced polarization may be responsible for a variety of interesting and anomalous electromechanical phenomena in materials including electromechanical coupling effects in non-uniformly strained nanostructures, "dead layer" effects in nanocapacitor systems, and "giant" piezoelectricity in perovskite nanostructures among others. In this work, adopting a lattice dynamics based microscopic approach we provide estimates of the flexoelectric tensor for certain cubic ionic crystals, perovskite dielectrics, III-V and II-VI semiconductors. We compare our estimates with experimental/theoretical values wherever available, address the discrepancy that exists between different experimental estimates and also re-visit the validity of an existing empirical scaling relationship for the magnitude of flexoelectric coefficients in terms of material parameters.

**1. Introduction**

In a continuum framework, the *linear* polarization response **P** to a strain field **ε** in a crystalline dielectric is typically given as

$$P_i = e_{ijk}\varepsilon_{ij} \qquad (1)$$

**e** is the third-rank piezoelectric tensor which couples strain to polarization. **e**, being an odd order tensor, vanishes identically for centrosymmetric crystals and thus only those dielectrics which possess a non-centrosymmetric crystal structure exhibit piezoelectricity.

In crystalline centrosymmetric dielectrics, where piezoelectricity is absent (**e**=0), a non-uniform strain can locally break the inversion symmetry of the unit cell, resulting in an induced dipole moment. In such a case, the bulk contribution to the polarization as a response to an applied macroscopic strain-gradient may be written as

---

♣ Corresponding author: psharma@uh.edu

$$P_i = \mu_{ijkl} u_{j,kl} \tag{2}$$

The phenomenological fourth-order tensor **μ** introduced in Eqn. (2) is known as the flexoelectric tensor and the associated phenomenon wherein a macroscopic strain gradient[1] induces a linear polarization response in a dielectric is termed *flexoelectricity* [1]. **μ**, being a tensor of even order, is non-zero for crystals of any symmetry. Therefore the polarization response to an applied deformation in a dielectric may be re-written as

$$P_i = e_{ijk}\varepsilon_{jk} + \mu_{ijkl} u_{j,kl} \tag{3}$$

The phenomenon of flexoelectricity in crystalline dielectrics was first predicted by Maskevich and Tolpygo [2]; a phenomenological description was later proposed by Kogan [3] who included a term coupling the polarization and the strain-gradient in the thermodynamic potential of the form

$$f_{ijkl} P_i u_{j,kl} \tag{4}$$

More recently, Tagantsev [4-5] has investigated this phenomenon in detail and has clarified several issues regarding the bulk nature of flexoelectricity and contributions due to surface and dynamic effects. The fourth order tensor **f** introduced in Eqn. (4) can be related to the flexoelectric tensor **μ** in Eqn. (3) and it symmetries are now well-known. Kogan [3] estimated the flexoelectric constants $\mu_{ijkl}$ to be of the order of *e/a*, where *e* is the electronic charge and *a*, the lattice parameter. Multiplication by the dielectric constant was later suggested which appears to have been confirmed experimentally in a series of studies by Cross and co-workers [6-9].

Yet another body of work, which parallels the theory of flexoelectricity in some ways, is the polarization gradient theory due to Mindlin [10-11]. Based on the long-wavelength limit of the shell-model of lattice dynamics, Mindlin [10] found that the core-shell and the shell-shell interactions could be incorporated phenomenologically by including the coupling of polarization gradients to strain and the coupling of polarization-gradients to polarization-gradients respectively in the thermodynamic potential (Eqns.5a-b)

$$\begin{gathered} d_{ijkl} P_{i,j} \varepsilon_{kl} \\ b_{ijkl} P_{i,j} P_{k,l} \end{gathered} \tag{5a-b}$$

The polarization-gradient strain coupling and the polarization strain-gradient coupling is often included in the energy density expression as a Lifshitz invariant [3,13] as shown in expression (6) on account of the fact that total derivatives cannot occur in the expression for energy.

---

[1] A macroscopic strain gradient implies that the gradient in strain exists over a macroscopically large length scale *L>>a* (where *a* is the characteristic length scale of the material; in crystals it can be chosen to be equal to the lattice parameter).

$$h_{ijkl}\left(u_{ij}P_{k,l} - P_k u_{ij,l}\right)$$ (6)

It can also be shown that the dispersive contributions due to the term (5a) in the thermodynamic potential involving polarization gradient terms and due to expression (4) involving flexoelectricity are of the same order in the wave vector and cannot be isolated from one another. The symmetries of the tensors **d** and **b** are also known [10]. Under the framework of Mindlin's polarization gradient theory, Askar et al [14] have arrived at numerical estimates of tensors **d** and **b** by relating them to shell model parameters for the cases of NaCl, NaI, KI, and KCl but as will be shown later, Askar et al's [14] estimates for the components of tensor **d** are more likely the values of a combination of components of tensor **d** and those of the tensor **f** which occurs in the context of flexoelectricity (Eqn.4).

In addition to the arguments presented above, yet another motivation to include higher order gradients of strain and polarization in the formulation of a continuum theory for crystalline dielectrics appears while investigating dynamic phenomena. Classical electromagnetism may be safely applied to excitations belonging to any part of the spectrum whereas *classical* linear elasticity (wherein the elastic energy involves only the first derivatives of displacement) is a "long wavelength theory" and designed to be applicable only in a certain frequency regime. Therefore a hybrid electromechanical theory is limited in its applicability due to its elastic part. The inclusion of gradients of strain and polarization along with higher-order inertia terms to the elastic part of the free energy can extend the applicability of a hybrid electromechanical field theory to frequencies in the region of 1 THz (far-infrared region) where dispersive effects become significant [15-16]. It should be noted that while the flexoelectric effect introduces spatial dispersion, polarization-gradients (and polarization-inertia effects) can model frequency dispersion effects.

The phenomenon of flexoelectricity in crystalline dielectrics [2] has been experimentally observed in a variety of contexts: bending of crystal plates [17] and measurements of thin films [18]. It has also been variously invoked to explain the anomalous capacitance of thin dielectric films [18] and the weak size-dependent piezoelectric behavior of carbon and boron-nitride nanotubes [19-20]. Macroscopic electromechanical effects in dislocated diatomic crystals of non-piezoelectric dielectrics, wherein large strain-gradients in the vicinities of dislocations lead to induced polarization [21], may also be explained using flexoelectricity. Some works have reported large flexoelectric effects in low dimensional systems such as nanographitic systems [22] and two dimensional boron-nitride sheets [23]. In addition, some recent theoretical works seem to suggest that flexoelectric effects can assume importance in various nanoscale electromechanical phenomena, especially in high-dielectric materials e.g. "giant" piezoelectricity in perovskite dielectric nanostructures, piezoelectric composites without using piezoelectric materials among others [24-26]. However, very few atomistic investigations to estimate the flexoelectric constants exist in the literature. Experimental determination of flexoelectric constants for some perovskite dielectrics have been carried out by Cross and co-workers [6-9] and Zubko et al [27] while from a theoretical viewpoint, Sahin and Dost [28] provide some estimates for $KTaO_3$ predicated on phonon dispersion data. In the present work, using an approach outlined by Tagantsev [4-5], we employ a lattice dynamics based method to extract the flexoelectric coefficients for certain representative ionic salts NaCl and KCl, III-IV semiconductors GaAs and GaP, II-VI semiconductors ZnO and ZnS, and finally high dielectric constant perovskites $BaTiO_3$(BTO), $SrTiO_3$(STO) and $PbTiO_3$ (PTO) in their cubic phases. We report estimates for the flexoelectric constants from both density functional theory (DFT) based *ab initio* lattice dynamics and empirical shell models. Wherever possible, we compare our results with previously published theoretical calculations or experimental results. Flexoelectric coefficients of perovskite dielectric materials are of particular interest---large flexoelectric effects have been consistently observed in experimental studies on bent thin films of high-permittivity perovskite dielectric materials [6-9] as well as atomistic simulations on bent nanostructures [22-23]. This has important ramifications in perovskite dielectric thin film/nanostructure based technologies such as nanocapacitors and energy harvesting applications [24-26, 29].

---

[2] It is interesting to note that the term flexoelectricity originated in the liquid crystal and biological membrane literature to describe curvature induced polarization in *flexed* membranes of orientable molecules. In this work however, we concern ourselves with flexoelectricity in crystalline dielectrics only.

The outline of our paper is as follows. In Section 2, we present a brief overview of a continuum theory involving the first-gradients of strain and polarization. We show how inclusion of appropriate terms in the electro-elastic energy density can lead to a linear polarization response to an applied strain gradient i.e. flexoelectricity. In Section 3, a microscopic lattice dynamics based analysis is carried out which identifies the atomistic origins of flexoelectricity. Certain subtleties associated with this phenomenon are also discussed. Section 4 outlines a recipe (due to Tagantsev [4-5]) to calculate bulk flexoelectric constants for crystalline dielectrics from lattice dynamical data. In Section 5, we bring out some differences between the approach of Tagantsev to calculate flexoelectric constants and that of Askar et al's [14] to calculate Mindlin's polarization gradient constants. The numerical values of the flexoelectric constants for some selected materials presented in Section 6. Finally we discuss the physical reasons responsible for the high flexoelectric constants displayed by perovskite dielectric materials in Section 7 as well as the reasons for the observed discrepancies between our theoretical estimates the available limited experimentally data.

## 2. Continuum flexoelectricity: Linear polarization response due to a strain gradient

The general formulation of an electromechanical theory involving first-gradients of strain and polarization has been discussed elsewhere [28]. Here we provide a brief summary. If one includes terms involving gradients of strain and polarization in the thermodynamic potential, then a hybrid internal energy density function can be written of the form

$$\Sigma = \frac{1}{2}a_{kl}P_kP_l + h_{ijk}P_iP_{j,k} + e_{ijk}P_i\varepsilon_{jk} + \frac{1}{2}b_{ijkl}P_{i,j}P_{k,l} + \frac{1}{2}c_{ijkl}\varepsilon_{ij}\varepsilon_{kl} \\ ... + d_{ijkl}P_{i,j}\varepsilon_{kl} + f_{ijkl}P_iu_{j,kl} + r_{ijklm}\varepsilon_{ij}u_{k,lm} + g_{ijklmn}u_{i,jk}u_{l,mn}... \tag{7}$$

**a**, **e** and **c** are the familiar second order reciprocal dielectric susceptibility tensor, third order piezoelectric tensor and the fourth order elastic constant tensor respectively. **f** is the fourth-order flexoelectric tensor introduced in (4) while **b** and **d** are fourth-order tensors from (5). The third order tensor **h** couples the polarization to its gradient while the fifth order tensor **r** couples strain and strain-gradient. Tensor **r** is sometimes referred to as the acoustic gyroscopic tensor. Tensor **g** represents elastic nonlocality and dictates the strength of the biquadratic strain-gradient coupling [15-16]: it also serves the purpose of smoothing out distribution of fields.

Balance equations and constitutive relations for the electromechanical stresses can be derived by carrying out a variational analysis of the Lagrangian derivable from Eqn. (7). The interested reader is referred to the paper by Sahin and Dost [28] wherein this variational analysis has been carried out in exhaustive detail.

In the absence of an external electric field and free charges, the following expression involving the polarization and its gradient can be deduced from the balance equations and constitutive laws

$$\left(a_{ij} + \varepsilon_0^{-1}\delta_{ij}\right)P_j = d_{ijkl}\varepsilon_{kl,j} - \left(e_{ijk}\varepsilon_{jk} + f_{ijkl}u_{j,kl}\right) + h_{ijk}\left(P_{k,j} - P_{j,k}\right) + b_{ijkl}P_{k,lj} \tag{8}$$

For a centrosymmetric material, the third order tensors in Eqn. (8) vanish

$$\left(a_{ij} + \varepsilon_0^{-1}\delta_{ij}\right)P_j = d_{ijkl}\varepsilon_{kl,j} - f_{ijkl}u_{j,kl} + b_{ijkl}P_{k,lj} \tag{9}$$

*The above expression shows that following the energy density expression of Eqn. (7), the polarization response is linearly related to the strain-gradient.*

From a microscopic point of view, the terms involving polarization-gradients in the expression for the internal energy density (7) can be shown to bear analogs to certain interaction energy terms occurring in a shell type lattice dynamical model. In particular, the shell-shell interactions can be modeled through the biquadratic coupling of polarization-gradients to themselves while the core-shell interactions can be modeled via the coupling of polarization-gradients to strain.

Using this approach, Askar *et al* [14] have carried out explicit calculations to estimate the independent components of the tensors **b** and **d** for NaCl and KCl in terms of corresponding shell model parameters. On the other hand, as discussed by Tagantsev [4-5], a simple rigid ion model, which approximates atoms as consisting of ionic cores devoid of a shell of electrons, suffices to make the connection with the phenomenological flexoelectric coupling. In the following section, we will outline Tagantsev's approach to calculating the flexoelectric constants using a simple rigid-ion model for lattice dynamics. Further, we will also bring out some important differences between Tagantsev's approach to capture flexoelectricity induced spatial dispersion using a rigid-ion model and Askar et al's [14] approach to capture polarization-gradient induced frequency dispersion using a shell-type lattice dynamical model. In doing so, we also hope to make physically transparent, the microscopic origins of both flexoelectricity and polarization-gradient effects.

### 3. Polarization due to a uniform strain gradient: Microscopic analysis

Following Tagantsev's description of the polarization response due to flexoelectricity [1], consider a *uniform* strain gradient in a macroscopically large (but finite) crystal

$$\varepsilon_{ij}(\mathbf{x}) = \varepsilon_{ij}(0) + \frac{\partial \varepsilon_{ij}}{\partial x_k} x_k \qquad (10)$$

Integrating both sides of Eqn. (10)

$$\int \varepsilon_{ij}(\mathbf{x}) d^3x = \int \varepsilon_{ij}(0) d^3x + \int \frac{\partial \varepsilon_{ij}}{\partial x_k} x_k d^3x \qquad (11)$$

If the gradient is uniform, then $\frac{\partial \varepsilon_{ij}}{\partial x_k}$ is constant and $\int x_k d^3x = 0$ (if one assumes that the crystalline structure under consideration is centered at the origin). Therefore,

$$\varepsilon_{ij}(0) = V^{-1} \int \varepsilon_{ij}(\mathbf{x}) d^3x \qquad (12)$$

Here $x_k$ are the Cartesian coordinates of a point inside the undeformed crystal.

In the presence of a strain given by Eqn. (12), a particle initially at **R** is shifted to position **R**'

$$\mathbf{R}' = \mathbf{R} + \mathbf{r} \qquad (13)$$

Here **r** is

$$r_i = \varepsilon_{ij}(0)R_j + \frac{1}{2}\frac{\partial \varepsilon_{ij}}{\partial x_k}R_j R_k + u_i^{(1)}(\mathbf{R}) + u_i^{(2)}(\mathbf{R}) \tag{14}$$

**u**$^{(1)}$(**R**) and **u**$^{(2)}$(**R**) are the linear response of the internal strain to the macroscopic strain $\varepsilon_{ij}$ and its gradient $\frac{\partial \varepsilon_{ij}}{\partial x_k}$. Following the assumption of linearity, **u**$^{(1)}$ and **u**$^{(2)}$ can be cast in the form

$$u_i^{(1)}(\mathbf{R}) = u_{i,p}^{(1)} = A_{i,p}^{jk}\varepsilon_{jk}(\mathbf{R}); \quad u_i^{(2)}(\mathbf{R}) = u_{i,p}^{(2)} = B_{i,p}^{jkl}\frac{\partial \varepsilon_{jk}}{\partial x_l}(\mathbf{R}) \tag{15}$$

$\mathbf{u}_p^{(1)}$ denotes the sub-lattice shift of the $p^{th}$ atom in the unit cell under the influence of a uniform strain; this quantity vanishes for all atoms in a centrosymmetric unit cell. On the other hand, $\mathbf{u}_p^{(2)}$ signifies the internal displacement of the $p^{th}$ atom in response to the applied strain-gradient and is non-zero in principle for crystals of any symmetry.

Following the displacements of expression (15), the polarization change due to such internal motions is given by

$$\delta \mathbf{P} = (V')^{-1}\sum_{\mathbf{R}'}Q(\mathbf{R}')\mathbf{R}' - (V)^{-1}\sum_{\mathbf{R}}Q(\mathbf{R})\mathbf{R} \tag{16}$$

$V$ and $V'$ are the volumes of the crystal before and after deformation and $Q(\mathbf{R})$ is the charge of the particle at **R**. From Eqns. (15-16),

$$\delta P_i = \underbrace{\varepsilon_{ij}(0)P_j^0 - \varepsilon_{jj}(0)P_i^0}_{\text{Spontaneous Polarization Contribution}} + \underbrace{(V)^{-1}\sum_{\mathbf{R}}Q(\mathbf{R})u_i^{(1)}(\mathbf{R})}_{\text{Piezoelectric Contribution}} + \\ + \underbrace{\frac{1}{6}Q_{jk}\frac{\partial \varepsilon_{ij}}{\partial x_k}}_{\text{Quadrupole moment contribution}} + \underbrace{\frac{I}{2}\frac{\partial \varepsilon_{ij}}{\partial x_k} + (V)^{-1}\sum_{\mathbf{R}}Q(\mathbf{R})u_i^{(2)}(\mathbf{R})}_{\text{Flexoelectric Contribution}} \tag{17}$$

In (17), $\mathbf{P}^0$ is the spontaneous polarization of the crystal in the undeformed configuration and **Q** is the average quadrupole moment density. $\mathbf{P}^0$, **Q** and **I** are defined as

$$\mathbf{P}^0 = (V)^{-1}\sum_{\mathbf{R}}Q(\mathbf{R})(\mathbf{R})$$
$$Q_{ij} = (V)^{-1}\sum_{\mathbf{R}}Q(\mathbf{R})(3R_iR_j - \delta_{ij}R^2) \tag{18a-c}$$
$$I = (V)^{-1}\sum_{\mathbf{R}}Q(\mathbf{R})R^2$$

As argued by Tagantsev [4], to estimate the flexoelectric response under an applied strain-gradient, the polarization induced should be measured under the conditions of zero macroscopic electric field ensuring the elimination of spurious spontaneous polarization and surface polarization effects [3]. Under such conditions, the spontaneous polarization $\mathbf{P}^0$ and the quadrupole moment density $\mathbf{Q}$ both vanish. Further, the induced polarization caused due to internal displacement of atoms $u_i^{(1)}(\mathbf{R})$ in response to a macroscopic strain corresponds to the piezoelectric effect. Thus, the induced polarization due to flexoelectricity can be isolated as

$$P_{fl} = \frac{I}{2}\frac{\partial \varepsilon_{ij}}{\partial x_k} + v^{-1} Q_p B_{i,p}^{jkl} \frac{\partial \varepsilon_{jk}}{\partial x_l} \qquad (19)$$

$v$ is the volume of the unit cell while $Q_p$ is the *effective* charge of the $p^{th}$ atom. The first term on the right hand side of (20) can be identified as the surface flexoelectric contribution [1-2] while the second term can be identified as the bulk flexoelectric contribution.

Thus the bulk flexoelectric tensor $\mu_{ijkl}$ can be identified from (19) as

$$\mu_{ijkl} = v^{-1} Q_p B_{i,p}^{jkl} \qquad (20)$$

It may be noted from (19) that polarization due to flexoelectricity is induced as a consequence of internal shifts among atoms within a unit cell due to an applied strain-gradient i.e. a dipole is created within a unit cell when atoms carrying opposite charges suffer a net displacement with respect to each other leading to a macroscopic polarization. Therefore, for flexoelectricity to exist, it is imperative that a strain gradient exists at the level of a unit cell i.e. there is a spatial variation of strain within the unit cell.

To further illuminate this point, consider the arrangement of atoms which form a part of a periodic 2-D ionic crystalline solid as shown in Fig (1). The red balls denote positive ions with unit charge while the blue ions denote negative ions with unit charge. We assume that this configuration is in stress-free equilibrium.

---

[3] The macroscopic electric field is associated with the non-analyticity of the lattice dynamical matrix at near zero wave vectors. Therefore, while investigating flexoelectric coefficients using lattice dynamical methods, the non-analytical contribution to the dynamical matrix should be removed. This point will be further elaborated in Section III.

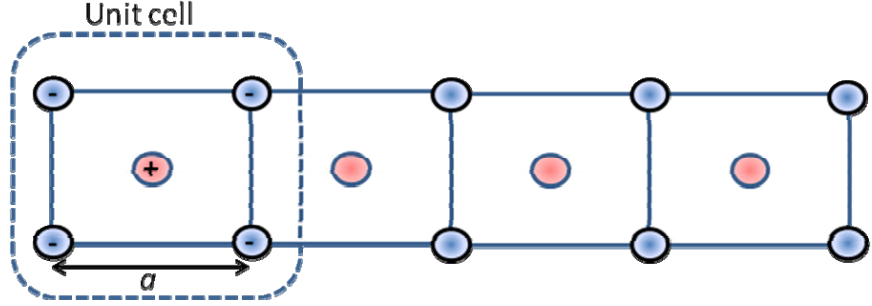

**Fig 1**: A finite undeformed arrangement of atoms in an ionic 2-D crystal. The unit cell chosen is highlighted within the dashed line. The blue balls denote anions while the red balls denote cations. Since the centers of positive and negative charge coincide, the net dipole moment is zero.

The dipole moment is given by

$$p = \frac{a}{2} - a + \frac{3a}{2} - 2a + \frac{5a}{2} - 3a + \frac{7a}{2} - \frac{4a}{2} = 0 \quad (21)$$

Thus, the dipole moment of the arrangement of atoms shown in Fig. (1) is zero.

Now we stretch each unit cell uniformly, but the amount of stretch in each unit cell is different such that an infinite strain gradient exists at the boundary of each unit cell (Fig.2).

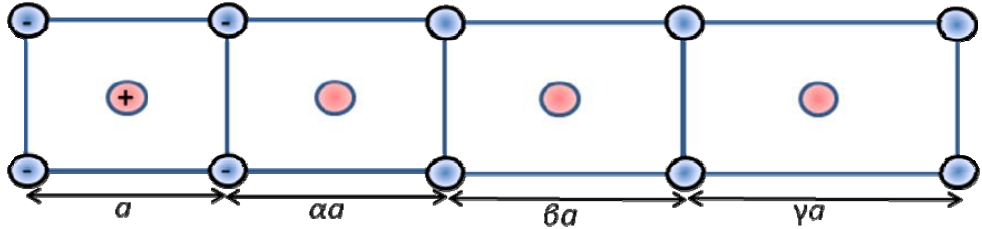

**Fig 2**: The configuration of Fig.1 is deformed in such a manner that each unit cell is stretched uniformly but the amount of stretch varies from cell to cell. Even though strain-gradients exist at the interfaces of the unit cells we have chosen, there is no net dipole moment created since the centers of positive charge and negative charge of each unit cell still coincide.

The dipole moment of the above arrangement becomes

$$p = \frac{a}{2} - a + \left(1 + \frac{\alpha}{2}\right)a - (1+\alpha)a + \left(1 + \alpha + \frac{\beta}{2}\right)a - (1+\alpha+\beta)a$$

$$... + \left(1 + \alpha + \beta + \frac{\gamma}{2}\right)a - (1+\alpha+\beta+\gamma)\frac{a}{2} = 0 \quad (22)$$

Thus even though strain-gradients exist at the interface of the unit cells, there is no induced polarization. For flexoelectricity to exist, the strain has to vary at the level of the unit cell such that the resulting internal shifts between differently charged atoms constituting the unit cell create dipoles. Further, for flexoelectricity to be a macroscopically observable effect, the strain gradient has to exist over a macroscopically large extent of the crystal such that the polarization averaged over several unit cells remains finite.

Yet another subtlety relates to the distinction between surface and bulk flexoelectricity. The phenomenon of flexoelectricity is pictorially explained as follows. Consider again an arrangement of atoms which form a part of periodic 2-D ionic crystalline solid as shown in Fig. (3a). For convenience, we only depict two unit cells.

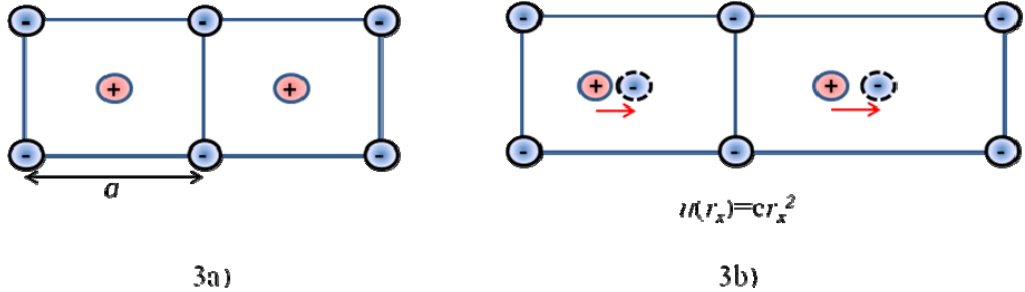

**Fig.3a**) Shows the undeformed stress free configuration of a portion of a 2-D diatomic ionic solid. **Fig. 3b**) Shows the deformed configuration wherein each atom is subjected to an inhomogeneous displacement of the form $u(r_x) = cr_x^2$, where $r_x$ is the x-coordinate of the position vector of that atom.

In the equilibrium stress-free configuration of Fig. (3a), the centers of positive and negative charges coincide and there is no net dipole moment. Now if an inhomogeneous displacement of the form $u(r_x) = cr_x^2$, where $r_x$ stands for the x-coordinate of an atom and $c$ is a constant, is applied to the stress-free configuration of Fig. (3a) and the atoms are clamped, then dipoles are created in each unit cell as is shown in Fig. (3b). However, this induced polarization is a result of surface flexoelectricity and corresponds to the first term on the right-hand side of Eqn. (19). If, under the conditions of an inhomogeneous stress, the atoms are 'unclamped' and allowed to relax, they undergo further internal shifts corresponding to the displacements $\mathbf{u}_p^{(2)}$ of Eqn. (15). It is the additional polarization created due to these internal shifts which corresponds to the bulk flexoelectric effect corresponding to the second term on the right-hand side of Eqn. (19).

Another point deserves mention. Even though, flexoelectricity in principle can be observed in all materials, one can see from the discussion above that in materials where effective charges of atoms $Q_p$ are zero, say for example a single element material like graphene where no effective charges can be assigned to atoms in the unit cell, the bulk flexoelectric constant of expression (20) becomes zero. However, flexoelectricity can still occur in such materials purely due to electronic wave function overlap effects. Indeed, Dumitrica et al. [18] have demonstrated the presence of flexoelectricity induced polarization in curved carbon nanoshells (Fig.4). The rigid-ion model can however not take into account such effects and this is indeed a limitation of the approach we adopt in this paper. In the following section we will outline an approach to calculate **B** (Eqn. 15) and subsequently the flexoelectric constant (Eqn. 20) from harmonic lattice dynamics.

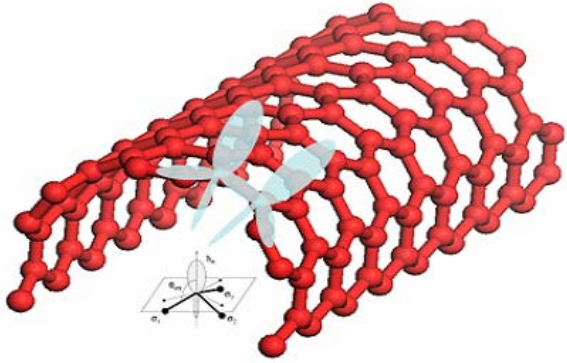

**Fig. 4** The curvature of a bent sheet of graphene results in rehybridization of p-orbital. As a result, the center of electronic charge at each atomic site is displaced outwards from the nuclear charge. This results in a curvature induced dipole moment.

### 4. Determination of flexoelectric constants: A lattice dynamics approach

Consider an acoustic wave traveling in an effectively infinite crystal with wave-vector **k** such that $|\mathbf{k}|^{-1}$ is much less than the crystal dimensions but much larger than the lattice parameter $a$. The displacement of the $p^{th}$ atom in the $n^{th}$ unit cell associated with such a wave can be written as

$$r_{i,p}^n = u_{i,p} e^{i(\mathbf{k}.\mathbf{R}_p^n - \omega t)} \quad (23)$$

The amplitude of displacement $\mathbf{u}_p$ corresponding to the $p^{th}$ atom of a unit cell corrected to include first order spatial and frequency dispersion effects can be written from Eqn. (14) as

$$u_{i,p} = w_i + iA_{i,p}^{jk} w_j k_k - B_{i,p}^{jkl} w_j k_k k_l - G_{i,p}^j w_j \omega^2 \quad (24)$$

In (24), **w** is the amplitude of the *pure* acoustic wave. For a pure acoustic phonon mode (**k**->0), the amplitude of displacement $u_{i,p}$ is independent of $p$ i.e. all atoms oscillate with the same amplitude of vibration. Physically speaking, this corresponds to a uniform deformation in classical continuum elasticity. The remaining terms on the right hand side of (24) correspond to internal shifts which occur because of the inherent discreteness of the crystal. The operator **G** corresponds to frequency dispersion corrections to the displacement amplitude and can be shown to be related to polarization inertia effects.

Now consider the Hamiltonian of a crystal written in the harmonic approximation

$$H = \Phi_0 + \frac{1}{2}\sum_{nip} m_p \left(\dot{r}_{i,p}^n\right)^2 + \frac{1}{2}\sum_{\substack{nip\\n'i'p'}} \Phi_{ip,i'p'}^{nn'} r_{i,p}^n r_{i',p'}^{n'} \qquad (25)$$

Here, $\Phi_0$ is the static (equilibrium) potential energy of the crystal and $\Phi_{ip,i'p'}^{nn'}$ constitute the elements of the so-called force constant matrix. In particular,

$$\Phi_{ip,i'p'}^{nn'} = \left.\frac{\partial^2 \Phi}{\partial r_{i,p}^n \partial r_{i',p'}^{n'}}\right)_0 \qquad (26)$$

Here, $\Phi$ is the total potential energy of the crystal assumed to be some function of the instantaneous positions of all the atoms.

Now the equations of motion for the lattice can be derived as

$$m_p \ddot{r}_{i,p}^n = -\frac{\partial \Phi}{\partial r_{i,p}^n} = -\sum_{n'i'p'} \Phi_{ip,i'p'}^{nn'} r_{i',p'}^{n'} \qquad (27)$$

The equations of motion (27) form an infinite set of simultaneous linear differential equations. Their solution can be simplified by exploiting the periodicity of the lattice if we choose as a solution to (27) a function of the form

$$r_{i,p}^n = u_{i,p} e^{i(\mathbf{k}.\mathbf{R}_p^n - \omega t)} \qquad (28)$$

After substituting the expression for **r** from Eqn. (28) into Eqn. (27), one can arrive at

$$\omega_j^2(k) u_{i,p}(\mathbf{k}j) = \sum_{i'p'} C_{ip,i'p'}(\mathbf{k}) u_{i',p'}(\mathbf{k}j) \qquad (29)$$

**C** is related to the dynamical matrix and can be written in terms of the force constants as

$$C_{ip,i'p'}(\mathbf{k}) = \sum_{n'} \Phi_{ip,i'p'}^{nn'} e^{-i\mathbf{k}.\left(\mathbf{R}_p^n - \mathbf{R}_{p'}^{n'}\right)} \qquad (30)$$

The set of equations given by Eq. (30) can be solved in a perturbative manner for small **k** by the method of long waves. We will accordingly expand all the quantities appearing in (30) in powers of **k** upto second order.

$$\begin{aligned}
C_{ip,i'p'}(\mathbf{k}) &= C_{ip,i'p'}^{(0)} + \sum_j C_{ip,i'p'}^{(1)j} k_j + \frac{1}{2}\sum_{\gamma\lambda} C_{ip,i'p'}^{(2)jl} k_j k_l + ...\\
u_{i,p}(\mathbf{k}j) &= u_{i,p}^{(0)}(\mathbf{k}j) + u_{i,p}^{(1)}(\mathbf{k}j) + u_{i,p}^{(2)}(\mathbf{k}j) + ...\\
\omega_j(\mathbf{k}) &= \omega_j^{(1)}(\mathbf{k}) + \omega_j^{(2)}(\mathbf{k}) + ...
\end{aligned} \qquad (31\text{a-c})$$

In case of ionic materials, the perturbative expansion of Eqns. (31a-c) presents problems because even the lowest order term in the expansion of the dynamical matrix diverges because of long-range electrostatic forces. This is dealt with by separating the electrostatic field at a point into a local Lorentzian field plus a global macroscopic electric field. Further, the contribution of the macroscopic field to the dynamical matrix can be identified with the non-analytical terms of the dynamical matrix which cause divergent behavior at near zero wave vectors. The short range contributions to the dynamical matrix due to short range forces and the Lorentz field can then be treated in a perturbative manner. However, as has been previously pointed out, the flexoelectric coefficients, by definition, measure the polarization response under the application of a uniform strain gradient in the absence of a macroscopic electric field. Thus in case of both weakly polar materials (like GaAs) and highly polar materials (like BTO) we exclude the contribution of the macroscopic electric field while calculating the dynamical matrix in Eqn. (31). This contribution is likely to be small for less polar materials like GaAs while one expects a large contribution due to the macroscopic field in a highly polar solid like BTO.

The expansion coefficients in Eqn. (31a) are given by

$$C^{(0)}_{ip,i'p'} = \sum_{n'} \Phi^{nn'}_{ip,i'p'}$$
$$C^{(1)j}_{ip,i'p'} = \sum_{n'} \Phi^{nn'}_{ip,i'p'} \left(\mathbf{R}^n_p - \mathbf{R}^{n'}_{p'}\right)_j \quad (32\text{a-c})$$
$$C^{(2)jl}_{ip,i'p'}(k,k') = -\sum_{n'} \Phi^{nn'}_{ip,i'p'} \left(\mathbf{R}^n_p - \mathbf{R}^{n'}_{p'}\right)_j \left(\mathbf{R}^n_p - \mathbf{R}^{n'}_{p'}\right)_l$$

As discussed before, the force constants occurring in Eqns. (32a-c) are such that the macroscopic field contribution has been excluded.

On substituting Eqns. (32a-c) in Eqns. (31a-c), we have

$$C^{(0)}_{ip,i'p'} u^{(0)}_{i'p'} = 0$$
$$C^{(0)}_{ip,i'p'} u^{(1)}_{i'p'} = -ik_j C^{(1)j}_{ip,i'p'} u^{(0)}_{i'p'} \quad (33\text{a-c})$$
$$C^{(0)}_{ip,i'p'} u^{(2)}_{i'p'} = -ik_j C^{(1)j}_{ip,i'p'} u^{(1)}_{i'p'} - \frac{k_j k_l}{2} C^{(2)jl}_{ip,i'p'} u^{(0)}_{i'p'} + \omega^2 m_p u^{(0)}_{ip}$$

One can solve for $u^{(0)}_{ip}$, $u^{(1)}_{ip}$ and $u^{(2)}_{ip}$ to obtain

$$\mathbf{u}^{(0)}_p = \mathbf{w}$$
$$u^{(1)}_{ip} = -\sum_{p''} \Gamma_{ip,i'p'} C^{(1)j}_{ip,i''p''} ik_j w_{i''} \quad (34\text{a-c})$$
$$u^{(2)}_{ip} = \sum_{p''} \Gamma_{ip,i''p''} \left(\omega^2 \mu_p \delta_{pp''} \delta_{i'i''} - k_j k_l \tilde{T}^{jl}_{i'p',i''p''}\right) w_{i''}$$

In Eqns. (34a-c), **w** is any arbitrary vector in space. The matrix $\Gamma$ in Eqns. (34b-c) is the *inverse* of the singular matrix defined in a special way. For a unit cell containing $r$ atoms, $p$ varying from 0 to $r$-1, the $3r \times 3r$ matrix $\Gamma$ in (34b-c) is defined as

$$\Gamma_{ip,i'p'} = \Gamma^{(3r-3)}_{ip,i'p'} \quad p, p' \neq 0$$
$$= 0 \quad \text{otherwise} \tag{35}$$

Here, $\Gamma^{(3r-3)}$ is the inverse to the $(3r-3) \times (3r-3)$ matrix $C^{(0)}_{ip,i'p'}$ ($p,p'$=1,2,…,r-1).

Further, the following definitions hold for the matrix $\tilde{\mathbf{T}}$ introduced in Eqn. (34c).

$$\tilde{T}^{jl}_{ip,i'p'} = T^{jl}_{ip,i'p'} - \frac{\delta_{pp'}}{s} \sum_{p'',p'''} T^{jl}_{ip'',i'p'''}$$

$$T^{jl}_{ip,i'p'} = C^{(1)j}_{ip,i''p''} \Gamma_{i''p'',i'''p'''} C^{(1)l}_{i'''p''',i'p'} + \frac{1}{2} C^{(2)jl}_{ip,i'p'} \tag{36a-b}$$

From Eqns. (34a-c) and (17), we can conclude that

$$A^{lj}_{i,p} = -\sum_{p''} \Gamma_{ip,i'p'} C^{(1)j}_{i'p',lp''}$$

$$B^{jkl}_{i,p} = \sum_{p''} \Gamma_{ip,i'p'} \tilde{T}^{kl}_{i'p',jp''} \tag{37}$$

$$G^{j}_{i,p} = -\Gamma_{ip,jp'} \mu_{p'}$$

Thus we arrive at expressions for **A**, **B** and **G** in terms of matrices which can be related to the real-space interatomic force constants.

One can in principle generate the phonon dispersion over a sufficiently large grid of wave-vectors by *ab-initio* or empirical lattice dynamics and then do an inverse Fourier transform in order to generate the inter-atomic force constants $\Phi^{nn'}_{ip,i'p'}$ up to a given number of neighbors corresponding to a rigid-ion lattice dynamical model. The denser the grid of phonon wave-vectors, the larger is the distance of the farthest neighbor to an atom for which interatomic constants can be calculated. Therefore, for a material like BTO for which long-range inter-atomic forces become important, one would be better served by generating the phonon dispersions over a large grid of wave-vectors.

5. **Tagantsev's approach to estimate flexoelectric constants vs. Askar et al.'s approach to calculate polarization-gradient constant**

Tagantsev's [4] approach to calculating flexoelectric constants employs a simple rigid-ion model. The flexoelectric polarization in this approach stems from the fact that in the long wavelength limit, different atoms (which correspond to ionic cores) in the same unit cell move by different amounts which corresponds to first order dispersive corrections. If one revisits the expression for the amplitude of displacement of the $p^{th}$ atom in a unit cell in the long-wavelength limit, one notices that the dispersive correction term involving **A** corresponds to the internal displacement of the atom in response to a uniform strain $\mathbf{u}_p^{(1)}$, while the terms involving **B** and **G** correspond to the internal displacement of the atom in response to an applied strain-gradient $\mathbf{u}_p^{(2)}$.

$$u_{i,p} = w_i + iA_{i,p}^{jk}w_j k_k - B_{i,p}^{jkl}w_j k_k k_l - G_{i,p}^{j}w_j \omega^2 \qquad (38)$$

The flexoelectric polarization simply spawns from the dipole created within a unit cell due to internal displacements of various ionic cores within the unit cell.

$$P_{flexo,i} = v^{-1}Q_p u_{i,p}^{(2)} \qquad (39)$$

On the other hand, Askar et al [14] use a shell type model to extract the polarization gradient constants **b** and **d** for centrosymmetric crystals NaCl, NaI, KCl, and KI. In order to illustrate their approach, consider a NaCl like crystal with two atoms per unit cell. In a shell like model, the outermost electron shell is considered to be a rigid spherical "shell" which can move with respect to the massive ionic "core" which consists of the nucleus and the inner electron shells. The position of $p^{th}$ atom in the $n^{th}$ unit cell is denoted by $\mathbf{r}_p^n$

The charge of the $p^{th}$ atom is given by

$$Q_p = X_p + Y_p \qquad (40)$$

Where $X_p$ and $Y_p$ are the charges of the core and shell of the $p^{th}$ atom respectively. The constraint of neutrality implies

$$\sum_p Q_p = 0 \qquad (41)$$

For the shell model, the positions of both the core and the shell, before deformation are given by $\mathbf{r}_p^n$. Their positions after deformation are respectively,

$$\begin{aligned}\mathbf{R}_p^{n,1} &= \mathbf{r}_p^n + \mathbf{u}_p^n \\ \mathbf{R}_p^{n,2} &= \mathbf{r}_p^n + \mathbf{u}_p^n + \mathbf{w}_p^n\end{aligned} \qquad (42)$$

**u** is the displacement of the core and **w** is the displacement of the shell with respect to the core.

The fact that the core and shell of each ion/atom carry different charges and that they can be displaced with respect to each other implies that when an effective electric field acts on the core and on the shell, they will suffer a relative displacement inducing a dipole moment at the ion/atom location porportional to the electric field strength. The proportionality constant is given by the polarizability of the ion $\alpha_p$ which enters into the shell model as a parameter. At the same time, even in the absence of an effective field, when two ion cores are brought closer together, the equilibrium positions of the centers of the corresponding shells need not coincide with the position of the cores, so that a dipole moment is induced on each ion/atom which is proportional to the displacement of the core. Thus the deformability and polarizability of each ion is included in the shell model. While in the rigid ion model, the dipole moment induced due to an electric field is only due to movement of rigid ions, in case of a shell model, additional contributions to the dipole moment arise as a result of the polarizability of the ion and also as a contribution due to the redistribution of charge in the region of overlap between neighboring ions. This latter contribution exists even in the absence of the first and is present for materials such as graphene and silicon which are made up of atoms and not ions. This is perhaps one of the biggest disadvantages of using a rigid ion model.

Now, under the assumption of a rigid ion model, let us consider an acoustic wave in the crystal such that

$$\mathbf{u}_p^n = \mathbf{u}_p e^{i(\mathbf{k}\cdot\mathbf{r}_p^n - \omega t)}; \mathbf{w}_p^n = \mathbf{w}_p e^{i(\mathbf{k}\cdot\mathbf{r}_p^n - \omega t)} \tag{43}$$

In the long wavelength limit, Askar et al [14], assume that the amplitude of displacement of the cores are the same i.e. $\mathbf{u}_p$ does not depend upon $p$. They neglect any internal displacements amongst the atoms as a result of first-order dispersive effects at low k-wavevectors. Instead, they assume a one-ion polarizable model wherein only one shell corresponding to a highly polarizable atom is capable of displacing with respect to its core. Say for example, in the case of NaCl, Na being numbered 1 and Cl being numbered 2, Askar et al [14] approximate $\mathbf{w}_1 = 0$ owing to the low polarizability of Na compared to that of Cl. Thus, in the long wavelength limit one has

$$\begin{aligned}\mathbf{u}_1 &= \mathbf{u}_2 = \mathbf{u} \\ \mathbf{w}_1 &= 0; \mathbf{w}_2 = \mathbf{w}\end{aligned} \tag{44}$$

The dipole moment per unit cell (i.e. the polarization is)

$$\mathbf{P} = \frac{1}{v}\left[Q_1\mathbf{u}e^{i\mathbf{k}\cdot\mathbf{r}_1} + (Q_2\mathbf{u} + Y_2\mathbf{w})e^{i\mathbf{k}\cdot\mathbf{r}_2}\right] \square \frac{1}{v}Y_2\mathbf{w} \tag{45}$$

Thus the polarization is attributed entirely to the displacement of the shell of the highly polarizable atom. In this regard, the displacement of the atoms **u** and the polarization which is decided by **w,** become independent quantities. In the rigid ion model on the other hand, the polarization and the displacement of atoms are inherently related since it is the relative displacement of the atomic cores which causes a dipole moment to arise. So in the rigid-ion model which is devoid of shells, the approach of Askar et al [14] will yield zero values for the polarization gradient constants.

**6. Results**

In this section we present the values for the bulk flexoelectric constants for

i) III-IV semiconductors GaAs , GaP and II-VI semiconductor ZnS

ii) Alkali Halides NaCl and KCl

iii) Perovskite dielectrics BTO and STO in their paraelectric phase.

Wherever possible we have tried to employ both *ab initio* and empirical shell model lattice dynamics to estimate the values for the flexoelectric constants. However, in some cases only one of these techniques is used either due to lack of accurate shell model potentials (for empirical lattice dynamics) or reliable pseudopotentials (for carrying out DFT based lattice dynamics). *Ab initio* phonon dispersions of GaAs were calculated in the local density approximation (LDA) using a norm-conserving pseudopotential generated by Giannozzi et al. [30], following a scheme proposed by von Barth and Car. A kinetic energy cut-off of 25 Rydbergs (Ry) was chosen and 60 **q** points were used for the Brillouin zone (BZ) integration. An equilibrium lattice parameter of 5.612 Å as suggested by Giannozzi et al. [30] was chosen. The dynamical matrices were generated on an 8×8×8 **k**-point mesh, and an inverse Fourier transform was carried out to generate the real space interatomic force constants. *Ab initio* phonon dispersions of BTO were calculated in the local density approximation (LDA) using Vanderbilt ultrasoft pseudopotentials. A kinetic energy cut-off of 90 Ry was chosen and a Monkhorst 6×6×6 grid of **q** points were used for the Brillouin zone (BZ) integration. An equilibrium lattice parameter of 4.00 Å was used. The dynamical matrices were generated on an 8×8×8 **k**-point mesh. *Ab initio* phonon dispersions of STO were calculated in the local density approximation (LDA) using Vanderbilt ultrasoft pseudopotentials. A kinetic energy cut-off of 90 Ry was chosen and a Monkhorst 6×6×6 grid of **q** points were used for the Brillouin zone (BZ) integration. An equilibrium lattice parameter of 3.85 Å was used. The dynamical matrices were generated on an 8×8×8 **k**-point mesh, and an inverse Fourier transform.Parameters for the shell lattice-dynamical model for GaAs, GaP and ZnS have been taken from Kunc et al. [31-32]. Parameters for NaCl and KCl were taken from Askar et al [14].

**Semiconductors**

For the case of the three semiconductors GaAs, GaP and ZnS, no previous estimates for the flexoelectric constants exist. Our estimates for the flexoelectric constants are summarized in Table 1.

|  | $\mu_{11}$ ($10^{-13}$C/cm) | | $\mu_{12}$ ($10^{-13}$C/cm) | | $\mu_{44}$ ($10^{-13}$C/cm) | |
|---|---|---|---|---|---|---|
|  | *Ab initio* | Shell model | *Ab initio* | Shell model | *Ab initio* | Shell model |
| GaAs | 0.5144 | 0.8512 | -0.8376 | 0.5107 | 0.2645 | 0.1702 |
| GaP | -na- | 0.4653 | -na- | 0.3128 | -na- | -0.3443 |
| ZnS | -na- | -0.311 | -na- | -1.544 | -na- | -0.611 |

**Table 1**: Flexoelectric constants for cubic semiconductors GaAs, GaP and ZnS from shell model lattice dynamics

The comparison of piezoelectric constants obtained from ab initio and shell model lattice dynamics with experimental values is given in Table 2.

| | $e_{14}$ (C/m$^2$) | | |
|---|---|---|---|
| | *Ab initio* | Shell model | Experiment |
| GaAs | -0.1464 | -0.066 | -0.16 |
| GaP | -na- | -0.0744 | -0.1 |
| ZnS | -na- | -0.111 | -0.13 |

**Table 2**: Piezoelectric constants for cubic semiconductors GaAs, GaP and ZnS obtained from shell model lattice dynamics compared with existing experimental values

**Alkali Halides**

For the case of the NaCl and KCl, we have employed only empirical lattice dynamics to estimate the flexoelectric constants. Askar et al. [14] have provided theoretical estimates using a single-ion polarizable shell model employing a different approach than ours. Our estimates using a similar model compare well with Askar et al.'s [14] estimates (Table 3).

| | $\mu_{11}$ (10$^{-13}$C/cm) | | $\mu_{12}$ (10$^{-13}$C/cm) | | $\mu_{44}$ (10$^{-13}$C/cm) | |
|---|---|---|---|---|---|---|
| | Shell model | Askar et al [14] | Shell model | Askar et al [14] | Shell model | Askar et al [14] |
| NaCl | 0.412 | 0.423 | -0.122 | -0.119 | -0.212 | -0.230 |
| KCl | 0.403 | 0.411 | -0.122 | 0.120 | -0.228 | -0.231 |

**Table 3**: Flexoelectric constants for cubic alkali halides NaCl and KCl obtained by shell model lattice dynamics compared with theoretical estimates by Askar et al. [11]

**Perovskite Dielectrics**

For the case of the perovskite dielectrics STO and BTO we have employed only *ab intio* lattice dynamics to estimate the flexoelectric constants. Experimental estimates for the flexoelectric constants exist due to Zubko et al. [27] (for STO) and Ma and Cross [9] (for BTO) and they are compared with our estimates in Table 4. As one can see, our estimates for the flexoelectric constants of STO possess the same order of magnitude as those experimentally provided by Zubko et al [27]. On the other hand our estimate for $\mu_{12}$ of BTO is smaller than that estimated by Ma and Cross by *three orders of magnitude*. It is interesting to note that in a recent *ab initio* study, scientists in Cambridge employed an alternative approach to estimate the flexoelectric constants of ferroelectric BTO and found them to be of the same order of magnitude as our estimates. The same group however report numbers close to Ma and Cross's [9] results for BTO while adopting an experimental approach. The possible reasons for such discrepancies are discussed in Section 7.

|  | $\mu_{11}$ ($10^{-13}$C/cm) | | $\mu_{12}$ ($10^{-13}$C/cm) | | $\mu_{44}$ ($10^{-13}$C/cm) | |
| --- | --- | --- | --- | --- | --- | --- |
|  | *Ab initio* | Experiment | *Ab initio* | Experiment | *Ab initio* | Experiment |
| STO | -26.4 | 20 | -374.7 | 700 | -357.9 | 300 |
| BTO | 15.0 | -na- | -546.3 | $10^6$ | -190.4 | -na- |

**Table 4**: Flexoelectric constants for cubic perovskite materials STO and BTO from ab initio calculations compared with available experimental data (Zubko et al. [27] for STO and Ma and Cross [9] for BTO)

In a recent *ab initio* study, scientists at Cambridge [33] have demonstrated, by employing an entirely different approach, that the flexoelectric constants for perovskite dielectric BTO and Lead Titanate and paraelectric STO have flexoelectric constants in the range of 1nC/m which, atleast to an order of magnitude, agrees with our estimates.

## 7. Discussion and Summary

In light of the values obtained for the materials considered in the previous section, it is clear that flexoelectric constants of perovskite dielectrics like BTO, STO are larger than those of conventional dielectrics like III-IV semiconductors, II-VI semiconductors and ionic salts like NaCl by as much as **four** orders of magnitude. This peculiar property of incipient perovskite dielectrics can be attributed not only to their anomalously large born effective charges but also to the existence of strong coupling between the transverse acoustic modes and the soft transverse optic modes so characteristic of incipient perovskite dielectrics [34]. This coupling lends itself to strong spatial dispersive effects which results in large atomic displacement responses to a non-homogeneous mechanical stimulus. Consequently, the internal sub-lattice shifts for such perovskite dielectrics due to an applied strain gradient may be orders of magnitude higher than those exhibited by conventional dielectrics. The transverse acoustic mode in such materials is known to exhibit anomalously large dispersion even at small k-vectors which suggest that elastic non-local effects in them (characterized by the tensor **g** in Eqn. (3)) may also be much larger than conventional materials.

As already indicated in the previous section, our estimates for STO are in the same order of magnitude as observed experimentally by Zubko et al. [27]. A few important factors should be kept in mind while interpreting the results. Typically, experiments to measure flexoelectric constants employ finite dimensional cantilever beams (for dynamic measurements) or thin films (subjected to static bending experiments). Due to the finite dimensions of these structures, surface polarization effects may affect the values of the measured flexoelectric constants. We report the **bulk** flexoelectric constants: surface flexoelectricity is omnipresent in experiments on finite structures [which the experiments employ] and does not disappear in the absence of a macroscopic electric field. There are indications though that surface flexoelectric constants maybe several orders of magnitude lesser than bulk flexoelectric constants for high permittivity materials and may not affect the values of the bulk flexoelectric constants measured in experiments. Further, broken symmetry at the surface of such finite structures may cause surface piezoelectricity which may contribute to the experimentally measured polarization. In addition as Zubko et al [27] point out, recent works [35] have indicated surface regions which are 100$\mu$ms deep with local fluctuations of the ferroelastic phase transitions that may induce spontaneous flexoelectric polarization in addition to that resulting from inhomogeneous stress caused during bending experiments. In view of the above mentioned points, Zubko et al. [27] suggest that their measurements should be viewed as order of magnitude estimates. Considering the latter caveat, our results and those of Zubko et al. are in broad agreement. Further, from a theoretical point of view, the phonon dispersions from *ab initio* simulations are extremely sensitive to kinetic energy cut-offs and the size of the grid employed to do Brillouin zone integrations: the estimates of the unstable modes (which are important to capture the large flexoelectric constants of perovskite dielectrics) are therefore somewhat suspect though we have ensured that the inter-atomic force constants we use are sufficiently converged.

However, for the case of BTO, there is a large discrepancy between our estimates and the experimental results of Ma and Cross [9]. Another independent group of workers from Cambridge [33] have used both *ab initio* and experimental techniques to estimate the flexoelectric constants for BTO. It is interesting to note that while our estimates for BTO match those of their *ab initio* estimates, there exists a large discrepancy with their experimental results which are closer to those published by Ma and Cross [9]. The reason for this may be the extreme sensitivity of the soft optic mode to temperature in such perovskite dielectrics. At finite temperatures, at which experiments are performed, a large TA-TO coupling may exist which in turn can explain the rather high value of the flexoelectric constants consistently observed by Cross and co-workers [6-9] for several materials. Since our lattice dynamics calculations assume zero temperature, there is a possibility of the existence of such a large discrepancy.

The magnitude of the flexoelectric constants are known to scale as $f = \lambda \varepsilon e / a$, $\varepsilon$ being the relative permittivity of the dielectric and $\lambda$ being a dimensionless scaling factor. While from our results it is clear that the flexoelectric constants do scale with the dielectric constant: small flexoelectric constants are observed for conventional dielectrics with $\varepsilon$ of the order of 10 and large flexoelectric constants are observed for perovskite dielectrics (3-4 orders of magnitude larger than conventional dielectrics) whose relative permittivity is of the order of $10^3$, our results (as well as experiments by Zubko et al. [27]) suggest that the empirical scaling factor $\lambda$ may be of the order of $10^{-2}$ which is in contrast to Ma and Cross [9] who estimate to $\lambda \sim 1$

**Acknowledgements:**

Financial support from NSF NIRT Grant No. CMMI 0708096 (Clark Cooper) and NSF Grant No. CMMI 826153 (Ken Chong) is gratefully acknowledged.